\documentclass[letterpaper]{article} % DO NOT CHANGE THIS
\usepackage[preprint]{aaai2027}  % DO NOT CHANGE THIS
\usepackage[hyphens]{url}  % DO NOT CHANGE THIS
\usepackage{graphicx} % DO NOT CHANGE THIS
\usepackage{natbib}  % DO NOT CHANGE THIS AND DO NOT ADD ANY OPTIONS TO IT
\usepackage{caption} % DO NOT CHANGE THIS AND DO NOT ADD ANY OPTIONS TO IT
\usepackage{algorithm}
\usepackage{algorithmic}

\usepackage{newfloat}
\usepackage{listings}
\DeclareCaptionStyle{ruled}{labelfont=normalfont,labelsep=colon,strut=off} % DO NOT CHANGE THIS
\floatstyle{ruled}
\newfloat{listing}{tb}{lst}{}
\floatname{listing}{Listing}

\usepackage{booktabs}

\usepackage{amsmath}
\usepackage{multirow}
\usepackage{seqsplit}

\title{MolReps: Taking SELFIES, Making SMILES, Teaching LLMs}
\author{
    Ken Deng, 
    Garrett Morris
}
\affiliations{
    Department of Statistics \\
    University of Oxford \\
    Oxford, United Kingdom \\
    ken.deng@stcatz.ox.ac.uk,
    garrett.morris@stats.ox.ac.uk
}

\begin{document}

\maketitle

\begin{abstract} 
The same molecule can be expressed in multiple, syntactically distinct representations, yet it remains unclear whether large language models (LLMs) can recognize and relate these equivalent encodings. We introduce \textbf{MolReps}, a pipeline that generates 9,600 question--answer pairs across ten representations for conversion, normalization, and 2D structure recognition. Post-training on MolReps improves text-only and multimodal models, \textbf{MolQwen3-4B-Instruct (MQ3-4B-I)} and \textbf{MolQwen3-VL-4B-Instruct (MQ3-VL-4B-I)}, from $4.01\%$ to $75.13\%$ and from $7.11\%$ to $73.14\%$ (InChIKey match) on the molecule-overlap test set, respectively. On the molecule-disjoint test set, post-training increased validity from 20.35\% to 58.12\% for MQ3-4B-I and from 17.19\% to 59.34\% for MQ3-VL-4B-I. However, InChIKey match remained low, decreasing from 3.86\% to 3.09\% and from 6.44\% to 2.29\%, respectively. Furthermore, the resulting models achieve a relative $27.84\%$ improvement in average over their base model on MolRepBench, a comprehensive benchmark for evaluating molecular property prediction. These results show the effectiveness of targeted post-training for enhancing molecular representation understanding and its potential to improve downstream molecular tasks.
% \footnote{Code will be available when published.}
\end{abstract}

% Uncomment the following to link to your code, datasets, an extended version or similar.
% You must keep this block between (not within) the abstract and the main body of the paper.
% Make sure that you do not de-anonymize yourself with these links.
\begin{links}
    % \link{Code}{https://aaai.org/example/code}
    \link{Code}{https://github.com/kdeng03/MolReps}
    % \link{Datasets}{https://aaai.org/example/datasets}
    \link{Datasets}{https://huggingface.co/datasets/kdeng03/mol-reps-v0.1}
    % \link{Extended version}{https://aaai.org/example/extended-version}
\end{links}

\section{Introduction}

Large language models (LLMs) have demonstrated strong capabilities across a wide range of applications~\citep{yang2025qwen3technicalreport,anthropic2026claude,comanici2025gemini,singh2025gpt5}. In chemistry, they are increasingly used for molecular property prediction~\citep{cavanagh2026smileyllamamodifyinglargelanguage}, drug discovery~\citep{li2025drugpilot}, and reaction planning~\citep{runcie2026chemiq}. Through pre-training on web-scale corpora, these models acquire substantial chemical knowledge~\citep{xia2025naturelanguagemodeldeciphering}, which can then be adapted to domain-specific tasks through post-training~\citep{li2025molr1explicitlongcotreasoning}. As molecular LLMs become more prevalent, however, a fundamental question remains: \textbf{How well do LLMs understand the different representations of the same molecule?}

Molecules are inherently structural objects, but computational systems require machine-readable encodings~\citep{bohacek1996art}. A single molecule may be expressed as SMILES~\citep{weininger1988smiles}, SELFIES~\citep{krenn2020self}, an InChI identifier~\citep{heller2015inchi}, an IUPAC name~\citep{favre2013nomenclature}, or another representation. Each format captures molecular structure through distinct syntactic and semantic conventions. They therefore differ in information density, human readability, and grammatical robustness. For example, SMILES is compact but may yield invalid strings, whereas SELFIES guarantees valid molecular strings at the cost of greater verbosity. These differences create a particular challenge for LLMs: although equivalent representations describe the same molecule, their sequential forms can differ substantially, making it difficult for a model to align their structural meaning~\citep{flam2022language}.

Recent work has shown that representation choice significantly affects downstream performance~\citep{baker2025molecular}. Nevertheless, existing molecular LLMs commonly use a single representation or a fixed set of formats~\citep{liu2024moleculargptopenlargelanguage,runcie2026chemiq}, without systematically assessing a model's understanding of each one. Instruction-tuned models such as LlaSMol~\citep{yu2024llasmoladvancinglargelanguage} and ChemLLM~\citep{zhang2024chemllmchemicallargelanguage} improve downstream task performance, but do not isolate representation understanding as a distinct capability. Similarly, \citet{raja2026rethinking} compares task outcomes across representations but does not directly measure an LLM's familiarity with each format. This gap limits principled decisions about which molecular representation is most suitable for a given model and task.

\begin{figure*}[t]
    \centering
    \includegraphics[width=\textwidth]{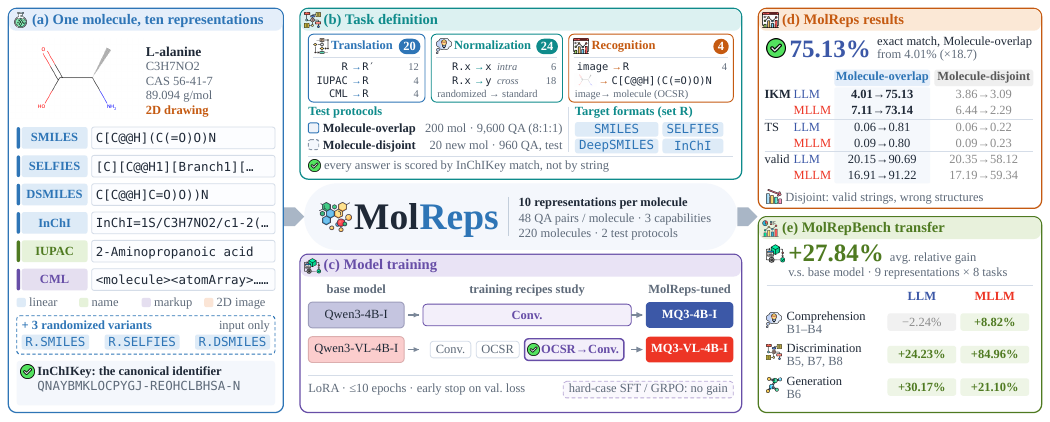}
    \caption{Overview of MolReps. (a) Ten representations of one
    molecule. (b) Six task templates, scored by InChIKey match under
    molecule-overlap and molecule-disjoint protocols. (c) LoRA
    post-training of text-only and multimodal Qwen3 models. (d)
    InChIKey match rises from 4.01\% to 75.13\% on molecule-overlap.
    (e) 27.84\% average gain on MolRepBench.}
    \label{fig:teaser}
\end{figure*}

To address this gap, we introduce \textbf{MolReps}, a dataset generation pipeline for measuring and improving molecular representation understanding (see Figure~\ref{fig:teaser}). MolReps is built on ChEBI-20-MM~\citep{liu2025chebi20mm}, a curated 20-class subset of the ChEBI database~\citep{hastings2015chebi}. It covers ten molecular representations and six task templates for textual conversion and 2D structure recognition. For each molecule, the pipeline generates 48 question--answer pairs that assess translation, normalization, and molecular-structure recognition across structurally equivalent encodings. Using 200 molecules, we generate 9,600 question--answer pairs in total.

Using MolReps, we formulate post-training as a \emph{recipe selection} problem. Inspired by success of introducing vision module in multimodal large language models (MLLMs)~\citep{liu2023llava, li2023blip2, wang2025internvl35}, we compare text-only and multimodal supervised fine-tuning recipes, including conversion-only training, 2D-recognition-only training, and sequential training from 2D recognition to conversion. The selected text-only and multimodal models, \textbf{MolQwen3-4B-Instruct} and \textbf{MolQwen3-VL-4B-Instruct}, improve from $4.01\%$ to $75.13\%$ and from $7.11\%$ to $73.14\%$, respectively. On the molecule-disjoint test set, post-training increased validity from 20.35\% to 58.12\% for MolQwen3-4B-Instruct and from 17.19\% to 59.34\% for MolQwen3-VL-4B-Instruct. However, InChIKey match remained low, decreasing from 3.86\% to 3.09\% and from 6.44\% to 2.29\%, respectively. In a second stage, we explore targeted refinement using rollout-derived hard cases through additional supervised learning and reinforcement learning. Under the current data scale and reward design, these refinements do not produce sufficiently consistent gains to replace the supervised checkpoints.

Finally, we evaluate the MolQwen3 family on MolRepBench~\citep{raja2026rethinking}, an external benchmark containing nine representations and eight chemical tasks, including atom counting, isomer discrimination, and caption-to-molecule generation. Across its primary comprehension, discrimination, and generation tasks, MolQwen3 improves on its base model by a relative $27.84\%$ in average. This result suggests that improved representation understanding has the potential to transfer to downstream molecular tasks.

Our contributions are as follows: \begin{enumerate}
    \item \textbf{MolReps}: A dataset generation pipeline spanning ten molecular representations and six task templates for evaluating representation conversion, normalization, and 2D structure recognition.
    \item \textbf{Post-training recipe study}: A controlled comparison of text-only and multimodal supervised post-training recipes that yields the MolQwen3 family, together with an exploratory study of rollout-based hard-case training that identifies its limitations as a reliable refinement method under the current data scale and reward design.
    \item \textbf{External evaluation}: An evaluation on MolRepBench of transfer beyond MolReps conversion tasks.
\end{enumerate}

\section{Related Work}

\paragraph{Molecular representations in cheminformatics.} Molecules can be expressed through a variety of encoding schemes, each with distinct inductive biases and practical trade-offs. SMILES~\citep{weininger1988smiles} remains the most widely used linear notation because it is compact and human-readable. However, its syntax can produce invalid strings and requires nontrivial parsing. Several alternatives have therefore been proposed. DeepSMILES~\citep{noel2018deepsmiles} simplifies the syntax for ring closures and branches, whereas SELFIES~\citep{krenn2020self,krenn2022selfies} uses a robust grammar that guarantees syntactic validity. InChI~\citep{heller2015inchi} provides a standardized, layered molecular identifier maintained by IUPAC, while IUPAC nomenclature~\citep{favre2013nomenclature} provides systematic chemical names. Beyond linear representations, structured formats such as Chemical Markup Language (CML)~\citep{murray1997chemical,murray1999chemical} and MolJSON~\citep{runcie2026molecularrepresentationslargelanguage} encode molecular graphs through XML and JSON schemas, respectively. This diversity enables different applications, but also raises the question of which formats language models can reliably interpret and manipulate.

\paragraph{Language models for molecular science.} Language models have been applied to a broad range of chemical and molecular tasks. A central task is \textbf{molecular property prediction}, including solubility (logS)~\citep{delaney2004esol}, lipophilicity (logP)~\citep{wildman1999prediction}, and toxicity~\citep{ames1975methods,ashby1988chemical,banerjee2018protox}. MoleculeNet~\citep{wu2018moleculenet} established widely used benchmarks for such tasks, and early sequence-based approaches used recurrent neural networks to learn molecular representations and predict molecular properties~\citep{segler2018generating}. ChemBERTa~\citep{chithrananda2020chembertalargescaleselfsupervisedpretraining} and MolBERT~\citep{fabian2020molbert} subsequently demonstrated the value of self-supervised language-model pre-training over molecular strings.

Another active direction connects molecular structures with natural language. Text2Mol~\citep{edwards2021text2mol} introduced text-to-molecule retrieval, and MolT5~\citep{edwards2022molt5} formulated molecular captioning and text-to-molecule generation within a unified sequence-to-sequence framework. More recent benchmarks, including ChemLLMBench~\citep{guo2023chemllmbench} and ChemIQ~\citep{runcie2026chemiq}, support systematic evaluation of chemical LLMs. Domain-specific models such as SmileyLlama~\citep{cavanagh2026smileyllamamodifyinglargelanguage} and NatureLM~\citep{xia2025naturelanguagemodeldeciphering} further adapt general-purpose LLMs to scientific and chemical data. Multimodal systems, including ChemVLM~\citep{li2025chemvlmexploringpowermultimodal}, Mol-LLM~\citep{lee2025molllmmultimodalgeneralistmolecular}, and Chem3DLLM~\citep{jiang2025chem3dllm3dmultimodallarge}, incorporate molecular graphs, images, or 3D structures alongside text. Mol-R1~\citep{li2025molr1explicitlongcotreasoning} explores reinforcement learning with rejection-sampled cold-start trajectories to improve molecular reasoning. Although these methods advance downstream task performance, they generally regard molecular representations as alternative input formats rather than investigating how well a model understands each representation.

\paragraph{Representation effects in molecular LLMs.} Recent evidence shows that the choice of molecular representation substantially affects LLM behavior. \citet{baker2025molecular} found that pretrained LLMs exhibit distinct preferences among molecular string representations in zero- and few-shot property prediction. MolRepBench~\citep{raja2026rethinking} further evaluates nine representations across eight tasks and shows that no single representation consistently dominates all downstream applications. These findings motivate representation-aware model design and evaluation.

Large chemical resources, including ChEBI~\citep{hastings2015chebi}, PubChem~\citep{kim2015pubchem}, ZINC250k~\citep{gomez2018automatic}, ChEMBL~\citep{zdrazil2024chembl}, and ChEBI-20-MM~\citep{liu2025chebi20mm}, have made it possible to construct diverse molecular training and evaluation data. Instruction-tuned molecular models, such as Mol-Instruction~\citep{fang2024molinstructions}, LlaSMol~\citep{yu2024llasmoladvancinglargelanguage}, ChemLLM~\citep{zhang2024chemllmchemicallargelanguage}, and MolecularGPT~\citep{liu2024moleculargptopenlargelanguage}, improve performance on molecular description, property prediction, chemical-domain adaptation, and in-context learning. However, prior work mainly evaluates aggregate task outcomes, rather than isolating representation understanding as a capability in its own right. Our work complements this literature by quantifying LLM familiarity with individual representations using targeted question--answer pairs from the same molecules, enabling finer-grained analysis of representation--model interactions.

\paragraph{Post-training for molecular LLMs.} Molecular LLMs are commonly adapted through supervised fine-tuning (SFT)~\citep{ouyang2022instructgpt,singh2025gpt5} and preference-based optimization, including RLHF, DPO, and GRPO~\citep{stiennon2020rlhf,rafailov2023dpo,shao2024deepseekmath,guo2025deepseekr1}. Existing instruction-tuning datasets include limited format-conversion examples, such as SMILES--IUPAC translation, alongside property prediction and molecular captioning~\citep{fang2024molinstructions,yu2024llasmoladvancinglargelanguage,zhang2024chemllmchemicallargelanguage}. However, they do not systematically examine how task composition, modality, or optimization method affects representation-focused performance. MolReps supports this comparison by using multi-representation conversion and normalization as explicit post-training signals.

\section{MolReps} \label{sec:molreps}

Figure~\ref{fig:teaser} gives an overview. In this section, we introduce \textbf{MolReps}, a molecular-representation benchmark that curates question-answer (QA) pairs for training and evaluating LLMs and MLLMs on cross-representation translation and normalization. Unlike prior molecular captioning benchmarks such as ChEBI-20, which target a single text-generation task, MolReps systematically covers translation among a diverse set of molecular representations.

\subsection{Data Curation} For each molecule, we collect ten complementary representations spanning four categories: \textbf{(1) linear notations}, including canonical and randomized SMILES, SELFIES, DeepSMILES, and InChI; \textbf{(2) structured markup}, namely Chemical Markup Language (CML), an XML-based encoding of the molecular graph; \textbf{(3) natural-language-like nomenclature}, namely the IUPAC name, which follows chemical nomenclature conventions but is not uniquely canonical; and \textbf{(4) visual representations}, namely 2D structural images. Canonical and randomized SMILES are generated from the molecular graph; SELFIES and DeepSMILES are then deterministically converted from their respective SMILES inputs, whereas InChI and CML are generated directly from the graph. Appendix~\ref{app:reps} lists the complete set of representations, their derivation, and the tools used for generation.

The selected representations vary both their surface syntax and the structure of the information they expose. SMILES, SELFIES, DeepSMILES, and InChI are all linear sequences, but they encode atom order, branches, rings, and molecular identity through different conventions. Randomized variants preserve the molecular graph while changing its traversal order. IUPAC names provide a compositional linguistic description, CML expresses a graph through hierarchical markup, and 2D images encode structure spatially rather than token by token. MolReps is therefore more than a collection of alternative strings: it tests whether a model can identify molecular information that remains invariant when the same graph is expressed in different forms.

\subsection{Task Definitions}\label{sec:tasks}
Each MolReps instance presents one representation of a molecule as input and requires the model to generate a target representation of the same underlying molecule. Because equivalent graphs admit many valid serializations, we do not score free-form output against the reference string. Instead we restrict the output space to four \textit{standardized target representations}:
\[
\mathcal{R}_{\text{std}} = \{\text{C.~{}SMILES},\ \text{SELFIES},\ \text{DeepSMILES},\ \text{InChI}\}.
\]
Each is produced by a fixed pipeline and therefore gives exactly one reference per molecule, and each has a stable comparison procedure: structure recovery is judged by the derived InChIKey rather than by string identity. Here C.~SMILES denotes the canonical form, as opposed to the randomized variant introduced below. This separates representation understanding from ambiguity in the evaluation target.

Two representations are collected but excluded from $\mathcal{R}_{\text{std}}$. CML is excluded because equivalent graphs serialize to multiple valid CML documents, and IUPAC names because several chemically valid names can denote the same structure; both would make a single reference arbitrary. SELFIES and DeepSMILES targets, by contrast, are derived deterministically from canonical SMILES. The randomized counterparts of the linear notations form the input-only set
\[
\mathcal{R}_{\text{rand}} = \{\text{R. SMILES},\ \text{R. SELFIES},\ \text{R. DeepSMILES}\},
\]
which is used exclusively as input for normalization tasks. Appendix~\ref{app:tasks} gives the six resulting task templates and their per-molecule counts.

The three capabilities below differ in whether the model receives a standardized or a randomized form and whether it must cross representation families.

\paragraph{Cross-representation translation.} The standardized representation translation task maps between any two distinct target representations ($r \in \mathcal{R}_{\text{std}} \to r' \in \mathcal{R}_{\text{std}}, r' \neq r$). It probes whether the model can translate through a shared encoding of molecular structure rather than relying solely on format-specific surface patterns. IUPAC-name understanding and CML understanding extend this setting to natural-language-like nomenclature and structured-markup inputs, respectively.

\paragraph{Normalization.} Intra-representation normalization maps a randomized input to the standardized target from the same representation family. Cross-representation normalization maps the same randomized input to the three standardized targets from the other representation families. Because the randomized input and its standardized counterpart describe the same graph through alternative traversals or serializations, normalization tests whether a model abstracts away from superficial string variation; the cross-representation variant additionally requires translation across families.

\paragraph{Multimodal understanding.} The 2D structure recognition task asks the model to read a molecular image, i.e., to perform optical chemical structure recognition (OCSR)~\citep{0903filippov2009optical,2304qian2023molscribe}, and to emit each standardized target for the depicted molecule. This evaluates whether the model can recover molecular-graph information from a diagram and encode it in machine-readable formats, a setting in which no input string is available to copy from.

\subsection{Data Statistics}
The source dataset is the validation set of ChEBI-20-MM~\citep{liu2025chebi20mm}, from which we sample 200 molecules. We sample from the validation split rather than the training split so that our source molecules are disjoint from those commonly used to train prior molecular LLMs, reducing the risk of inflating results with molecules the community has already overused. Although we only use a 200-molecule subset, we note that this is a pipeline and can be scaled to an unlimited MolReps-like dataset for future research.

Based on the task templates in Appendix~\ref{app:tasks}, each molecule yields $20 + 12N + 4$ QA pairs, where $N$ is the number of randomized SMILES drawn per molecule. To keep the dataset balanced, we set $N=2$, so each molecule contributes $20 + 24 + 4 = 48$ QA pairs. For the entire dataset, we create $200 \times 48 = 9600$ QA pairs; Appendix~\ref{app:stats} reports the resulting split sizes and the distribution of task types.

\paragraph{Data Splits.} We design two complementary evaluation protocols. The first is a \textit{molecule-overlap} (MO) protocol, which measures whether the model can acquire representation-translation skills for molecules observed during training. We partition the 9600 QA pairs into training, validation, and test sets with a 0.8/0.1/0.1 ratio, yielding 7680, 960, and 960 pairs. Because multiple QA pairs derive from the same molecule, this random split at the QA level may assign different representations of the same molecule to both training and test sets. We refer to this as \textit{molecule leakage}: with 48 pairs per molecule, virtually all 200 molecules appear in every split. Consequently, MO evaluates whether the model can learn the translation mapping for familiar molecules and supports recipe selection, but high MO performance can reflect molecule-specific learning rather than general representation rules.

The second is a \textit{molecule-disjoint} (MD) protocol, which measures whether this capability transfers to molecules never encountered during training. We sample an additional 20 molecules from the same source, disjoint from the 200 source molecules, and generate QA pairs for them using the same pipeline, yielding 960 extra test instances. Since a successful model must apply learned representation operations to unfamiliar graphs rather than retrieve a graph observed through another QA pair, MD is the stricter test. Reporting both protocols makes the distinction between within-molecule completion and structural generalization explicit; Appendix~\ref{app:stats} summarizes the resulting splits.

\begin{figure}[tb]
    \centering
    \includegraphics[width=\linewidth]{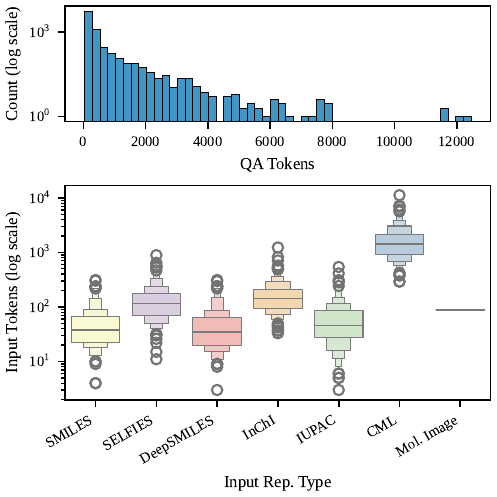}
    \caption{Token-count distributions in MolReps under the Qwen3 tokenizer. The top panel shows the total token count per QA instance; the bottom panel counts the number of tokens for input representations.}
    \label{fig:token_distribution}
\end{figure}

\paragraph{Token Counting.} Figure~\ref{fig:token_distribution} quantifies the sequence-length characteristics of MolReps under the Qwen3 tokenizer. Total QA lengths follow a right-skewed, long-tailed distribution: most instances are compact while a small number require substantially longer contexts. Among inputs, 2D structural images incur the greatest overhead, whereas SMILES, SELFIES, DeepSMILES, and IUPAC names are comparatively compact; among targets, InChI and SELFIES produce longer and more variable sequences than SMILES-based targets. Because the choice of input--output pair directly affects context length, memory use, and computational cost, this heterogeneity is relevant to both training and inference.

\section{Experimental Setup}

This section specifies the backbones, metrics, training configuration, and post-training recipes used throughout the paper. We treat post-training as a recipe selection problem: we fix a supervised recipe for each backbone and later test whether rollout-based refinement improves on it.

\paragraph{Backbones and data.} We use Qwen3-4B-Instruct~\citep{yang2025qwen3technicalreport} as the text-only backbone and Qwen3-VL-4B-Instruct~\citep{bai2025qwen3vltechnicalreport} as the multimodal backbone. The text-only backbone is evaluated only on textual tasks, whereas the multimodal backbone is evaluated on both textual conversion and 2D structure recognition. Both are trained on the 200-molecule MO split and evaluated on its held-out QA pairs and on the 20-molecule MD split, following the protocols of Section~\ref{sec:molreps}; Appendix~\ref{app:stats} lists the split sizes. Our source molecules come from ChEBI-20-MM~\citep{liu2025chebi20mm}, a curated 20-class subset of ChEBI~\citep{hastings2015chebi}.

\paragraph{Metrics.} We report validity, InChIKey Match (IKM), and Tanimoto similarity (TS). Validity is the proportion of outputs that can be parsed and chemically sanitized in the requested representation via \texttt{RDKit}. IKM is the proportion of valid predictions whose derived InChIKey matches the reference; it is our primary metric for exact structure recovery. TS is the mean Tanimoto similarity between Morgan fingerprints of predicted and reference molecules, using radius 2 and 2,048 bits, and provides a graded measure of structural similarity. Unless stated otherwise, we report \textit{point estimates and bootstrap confidence intervals} obtained from 1,000 resamples of evaluation instances.

\paragraph{Training configuration.} All post-training runs use LoRA adapters. Training runs for at most ten epochs, with early stopping after two consecutive epochs without validation-loss improvement. We save a checkpoint after each epoch and evaluate the checkpoint with the lowest validation loss. Appendix~\ref{app:training} summarizes the optimization configuration. Each experiment uses a fixed random seed and is trained on one NVIDIA RTX 4090 GPU with 24~GiB of memory. We use \texttt{SFTTrainer} and \texttt{GRPOTrainer} from \texttt{trl} for post-training, and \texttt{vLLM} for inference and GRPO rollouts. SFT runs require approximately six hours, whereas GRPO runs require approximately 30 hours.

\paragraph{Recipes.} The text-only recipe fine-tunes Qwen3-4B-Instruct on the textual portion of the MO training set, which contains the five text-input MolReps templates (standardized translation, IUPAC understanding, CML understanding, and both normalization variants); we denote the result \textbf{MQ3-4B-I}. For the multimodal backbone we compare three curricula. The conversion-only model, MQ3-VL-4B-I$_{\text{Conv.}}$, uses the same textual pairs as the text-only recipe. The 2D-recognition-only model, MQ3-VL-4B-I$_{\text{OCSR}}$, trains on the 640 MO pairs whose inputs are 2D structural diagrams. The sequential model, MQ3-VL-4B-I$_{\text{O.}\to\text{C.}}$, first trains on 2D structure recognition and then on textual conversion. These three checkpoints isolate the effect of textual, visual, and sequential training, and Section~\ref{sec:main} selects among them. We denote the final selected checkpoint as \textbf{MQ3-VL-4B-I}.

\paragraph{External benchmark.} We additionally evaluate the base models and their trained counterparts on MolRepBench~\citep{raja2026rethinking}, which covers eight molecular tasks across comprehension, discrimination, and generation; Table~\ref{tab:molrepbench_tasks} lists the tasks and their metrics. Several MolRepBench tasks draw on ChEBI-20, so we interpret this evaluation as external task transfer rather than a disjoint test set.

\section{Results}

We organize the results around four questions: whether post-training teaches cross-representation conversion on MolReps (Section~\ref{sec:main}), what MolReps reveals about task and representation difficulty (Section~\ref{sec:reveals}), whether refinement beyond supervised training helps (Section~\ref{sec:rollout}), and whether the gains transfer to external molecular tasks (Section~\ref{sec:transfer}).

\subsection{Main Results: MO versus MD}
\label{sec:main}

We compare the base models with the final trained checkpoints, MQ3-4B-I and MQ3-VL-4B-I, under both MolReps protocols. MO evaluates cross-representation completion for molecular graphs observed during training, whereas MD evaluates transfer to held-out molecular graphs.

\paragraph{Molecule-overlap performance.} Table~\ref{tab:main_results} shows substantial MO gains after post-training. On textual conversion, IKM rises from $4.01\%$ to $75.13\%$ for the LLM and from $7.11\%$ to $73.14\%$ for the sequential MLLM. Validity reaches $89\%$--$91\%$, and Tanimoto similarity reaches $0.78$--$0.81$. These results show that both models can learn cross-representation completion for molecular graphs represented in the training data. Their similar point estimates suggest that both text-only and multimodal backbones support this behavior under the MO protocol.

\begin{table}[tb]
    \centering
    \footnotesize
    \setlength{\tabcolsep}{2pt}

\newcommand{\CLB}{\cellcolor{color_light_blue}}
\newcommand{\CB}{\textcolor{color_blue}}
\newcommand{\CLR}{\cellcolor{color_light_red}}
\newcommand{\CO}{\textcolor{color_orange}}
\newcommand{\CG}{\textcolor{color_green}}
\newcommand{\CR}{\textcolor{color_red}}

\definecolor{best}{HTML}{FFF2CC}
\newcommand{\be}{\cellcolor{best}}
\definecolor{second}{HTML}{E7E6E6}
\newcommand{\se}{\cellcolor{second}}

\resizebox{\linewidth}{!}{
\begin{tabular}{ccl|ccc}
\toprule
 &  &  & Valid. (\%) & IKM (\%) & TS \\
% Test Subset & Task Type & Model Type & Model &  &  &  &  \\
\midrule
\multicolumn{3}{l}{\textit{Conversion Task (text)}} \\
\midrule
\multirow[c]{6}{*}{MO} & \multirow[c]{2}{*}{LLM} & Q3-4B-I & 20.15$_{\pm 1.31}$ & 4.01$_{\pm 0.67}$ & 0.06$_{\pm 0.01}$ \\
 &  & MQ3-4B-I & 90.69$_{\pm 1.00}$ & \textbf{75.13$_{\pm 1.50}$} & \textbf{0.81$_{\pm 0.01}$} \\
\cmidrule{2-3} 
 & \multirow[c]{4}{*}{MLLM} & Q3-VL-4B-I & 16.91$_{\pm 1.30}$ & 7.11$_{\pm 0.84}$ & 0.09$_{\pm 0.01}$ \\
 &  & MQ3-VL-4B-I$_{OCSR}$ & 25.50$_{\pm 1.51}$ & 5.01$_{\pm 0.78}$ & 0.09$_{\pm 0.01}$ \\
 &  & MQ3-VL-4B-I$_{Conv.}$ & 89.39$_{\pm 1.04}$ & 70.84$_{\pm 1.55}$ & 0.78$_{\pm 0.01}$ \\
 &  & MQ3-VL-4B-I & \textbf{91.22$_{\pm 0.95}$} & 73.14$_{\pm 1.51}$ & 0.80$_{\pm 0.01}$ \\
% \cmidrule{2-7} 
\midrule
\multirow[c]{6}{*}{MD} & \multirow[c]{2}{*}{LLM} & Q3-4B-I & 20.35$_{\pm 1.36}$ & 3.86$_{\pm 0.65}$ & 0.06$_{\pm 0.01}$ \\
 &  & MQ3-4B-I & 58.12$_{\pm 1.71}$ & 3.09$_{\pm 0.59}$ & 0.22$_{\pm 0.01}$ \\
\cmidrule{2-3}
 & \multirow[c]{4}{*}{MLLM} & Q3-VL-4B-I & 17.19$_{\pm 1.29}$ & \textbf{6.44$_{\pm 0.82}$} & 0.09$_{\pm 0.01}$ \\
 &  & MQ3-VL-4B-I$_{OCSR}$ & 29.78$_{\pm 1.56}$ & 4.79$_{\pm 0.70}$ & 0.10$_{\pm 0.01}$ \\
 &  & MQ3-VL-4B-I$_{Conv.}$ & \textbf{62.44$_{\pm 1.62}$} & 2.17$_{\pm 0.49}$ & \textbf{0.24$_{\pm 0.01}$} \\
 &  & MQ3-VL-4B-I & 59.34$_{\pm 1.69}$ & 2.29$_{\pm 0.52}$ & 0.23$_{\pm 0.01}$ \\
\midrule
\multicolumn{3}{l}{\textit{OCSR Task (vision)}} \\
\midrule
 \multirow[c]{4}{*}{MO} & \multirow[c]{4}{*}{MLLM} & Q3-VL-4B-I & 23.51$_{\pm 4.78}$ & 1.28$_{\pm 1.21}$ & 0.08$_{\pm 0.02}$ \\
 &  & MQ3-VL-4B-I$_{OCSR}$ & 43.61$_{\pm 5.46}$ & 3.91$_{\pm 2.08}$ & 0.18$_{\pm 0.03}$ \\
 &  & MQ3-VL-4B-I$_{Conv.}$ & 63.57$_{\pm 5.20}$ & 19.94$_{\pm 4.52}$ & 0.34$_{\pm 0.04}$ \\
 &  & MQ3-VL-4B-I & \textbf{68.55$_{\pm 5.09}$} & \textbf{26.06$_{\pm 4.96}$} & \textbf{0.41$_{\pm 0.05}$} \\
% \cmidrule{2-8}
% \cmidrule{2-7} 
\midrule
 \multirow[c]{4}{*}{MD} & \multirow[c]{4}{*}{MLLM} & Q3-VL-4B-I & 18.89$_{\pm 4.39}$ & 1.28$_{\pm 1.21}$ & 0.07$_{\pm 0.02}$ \\
 &  & MQ3-VL-4B-I$_{OCSR}$ & 41.24$_{\pm 5.50}$ & \textbf{3.74$_{\pm 2.17}$} & \textbf{0.18$_{\pm 0.03}$} \\
 &  & MQ3-VL-4B-I$_{Conv.}$ & \textbf{58.87$_{\pm 5.41}$} & 0.00$_{\pm 0.00}$ & 0.16$_{\pm 0.02}$ \\
 &  & MQ3-VL-4B-I & 55.05$_{\pm 5.47}$ & 1.28$_{\pm 1.20}$ & 0.17$_{\pm 0.03}$ \\
\bottomrule
\end{tabular}
}

    \caption{Overall MolReps results for base models and training recipes.
    MO denotes molecule-overlap evaluation and MD denotes molecule-disjoint
    evaluation. Valid.: validity; IKM: InChIKey Match; TS: Tanimoto similarity.}
    \label{tab:main_results}
\end{table}

\paragraph{Molecule-disjoint performance.} MD results reveal a substantial MO--MD gap. On textual conversion, IKM falls from $75.13\%$ on MO to $3.09\%$ on MD for the LLM and from $70.84\%$ to $2.17\%$ for the conversion-only MLLM. Thus, neither model reliably recovers the exact structures of held-out molecular graphs. Nevertheless, post-training improves MD validity and Tanimoto similarity. For example, the MLLM improves from $17.19\%$ to $62.44\%$ validity and from $0.09$ to $0.24$ TS, while the LLM improves from $20.35\%$ to $58.12\%$ validity and from $0.06$ to $0.22$ TS. Post-training therefore transfers some output-format control and approximate structural similarity, but not reliable exact structure recovery, to MD molecules.

\paragraph{Multimodal recipe comparison.} Table~\ref{tab:main_results} compares four MLLM recipes: the base VLM, OCSR-only post-training from image to text, conversion-only post-training from text to text, and sequential OCSR-to-conversion training. Training on one input modality provides limited transfer to the other. The conversion-only model reaches 70.84\% IKM on MO textual conversion but 19.94\% on MO 2D structure recognition. Conversely, OCSR-only post-training reaches 5.01\% on MO textual conversion and 3.91\% on MO 2D structure recognition. These results show that neither single-task strategy yields strong performance on both modalities.

The two-stage strategy, OCSR followed by conversion, yields the highest MO point estimates on both textual conversion (73.14\% IKM) and 2D structure recognition (26.06\% IKM). On MD textual conversion, however, OCSR-only post-training reaches 4.79\% IKM, compared with 2.17\% and 2.29\% for conversion-only and two-stage training, respectively. Exact MD recovery remains low for all configurations.

Overall, we select the sequential curriculum as the final multimodal model, denoted \textbf{MQ3-VL-4B-I}: it gives the highest 2D-recognition IKM ($26.06\%$) and the highest textual-conversion validity ($91.22\%$) while retaining $73.14\%$ textual IKM, and training on one modality alone transfers poorly to the other. MQ3-4B-I is the text-only counterpart. The remaining recipes are retained as ablations, and Appendix~\ref{app:curves} shows that training and validation loss decrease together for both, so likelihood-based checkpoint selection is well behaved; this does not by itself establish molecule-disjoint generalization. No evaluated strategy achieves reliable MD exact recovery.

\subsection{What MolReps Reveals}
\label{sec:reveals}

\paragraph{Task-level analysis.} Table~\ref{tab:conv_results} breaks the MO conversion results down by task type. Its rows separate the five text-input templates: representation translation, IUPAC understanding, CML understanding, intra-representation normalization, and cross-representation normalization, and its columns compare each base model with its trained counterpart on validity, IKM, and TS. 

Two findings stand out. First, the base models convert \emph{almost nothing}: only intra-representation normalization is non-zero, at $30.79\%$ IKM for the text-only model and $53.56\%$ for the multimodal one, while the remaining four categories stay near zero. Relating equivalent representations is therefore not a capability that pretraining supplies; it has to be taught. Second, post-training lifts every category \emph{but far from uniformly}. Normalization is easiest ($77\%$--$89\%$ IKM), followed by representation translation ($67\%$--$68\%$) and CML understanding ($56\%$--$57\%$), whereas IUPAC understanding is hardest, reaching only $36\%$--$38\%$ despite validity above $74\%$. The ordering tracks how much surface regularity a task preserves: normalization pairs reuse the syntax of their representation family and change only the traversal order, whereas IUPAC names restate the graph in a separate nomenclature system with no shared tokens. Cross-representation normalization in fact slightly exceeds the intra-representation variant ($86\%$--$89\%$ versus $77\%$--$81\%$), so difficulty is not set by whether input and output belong to the same family; what matters is whether the mapping can lean on shared surface form.

\begin{table}[tb]
    \centering
    \footnotesize
    % Set column width
    \setlength{\tabcolsep}{2pt}

\resizebox{\linewidth}{!}{
\begin{tabular}{lll|ccc}
\toprule
 &  &  & Valid. (\%) & IKM (\%) & TS \\
% Model Version & Model & Tasks &  &  &  \\
\midrule
\multirow[c]{10}{*}{\rotatebox{90}{Base}} & \multirow[c]{5}{*}{Q3-4B-I} & Rep. Trans. & 15.11$_{\pm 2.26}$ & 0.00$_{\pm 0.00}$ & 0.01$_{\pm 0.00}$ \\
 &  & IUPAC Und. & 27.34$_{\pm 5.23}$ & 0.00$_{\pm 0.00}$ & 0.02$_{\pm 0.01}$ \\
 &  & CML Und. & 11.86$_{\pm 3.66}$ & 0.00$_{\pm 0.00}$ & 0.00$_{\pm 0.00}$ \\
 &  & Intra Norm. & 49.12$_{\pm 4.71}$ & 30.79$_{\pm 4.46}$ & 0.39$_{\pm 0.04}$ \\
 &  & Cross Norm. & 15.21$_{\pm 1.84}$ & 0.00$_{\pm 0.00}$ & 0.01$_{\pm 0.00}$ \\
\cmidrule(lr){2-3}
 & \multirow[c]{5}{*}{Q3-VL-4B-I} & Rep. Trans. & 5.59$_{\pm 1.46}$ & 0.40$_{\pm 0.40}$ & 0.01$_{\pm 0.00}$ \\
 &  & IUPAC Und. & 13.06$_{\pm 3.99}$ & 0.00$_{\pm 0.00}$ & 0.02$_{\pm 0.01}$ \\
 &  & CML Und. & 16.69$_{\pm 4.22}$ & 0.00$_{\pm 0.00}$ & 0.01$_{\pm 0.00}$ \\
 &  & Intra Norm. & 70.08$_{\pm 4.36}$ & 53.56$_{\pm 4.85}$ & 0.61$_{\pm 0.04}$ \\
 &  & Cross Norm. & 8.68$_{\pm 1.39}$ & 0.00$_{\pm 0.00}$ & 0.01$_{\pm 0.00}$ \\
\midrule
\multirow[c]{10}{*}{\rotatebox{90}{Trained}} & \multirow[c]{5}{*}{MQ3-4B-I} & Rep. Trans. & 88.61$_{\pm 2.00}$ & 68.23$_{\pm 3.00}$ & 0.76$_{\pm 0.03}$ \\
 &  & IUPAC Und. & 74.04$_{\pm 5.42}$ & 37.60$_{\pm 5.79}$ & 0.49$_{\pm 0.05}$ \\
 &  & CML Und. & 77.18$_{\pm 4.95}$ & 56.89$_{\pm 5.70}$ & 0.65$_{\pm 0.05}$ \\
 &  & Intra Norm. & 98.27$_{\pm 1.25}$ & 80.78$_{\pm 3.65}$ & 0.88$_{\pm 0.03}$ \\
 &  & Cross Norm. & 95.60$_{\pm 1.06}$ & 88.61$_{\pm 1.69}$ & 0.92$_{\pm 0.01}$ \\
\cmidrule(lr){2-3}
 & \multirow[c]{5}{*}{MQ3-VL-4B-I} & Rep. Trans. & 91.21$_{\pm 1.78}$ & 67.54$_{\pm 2.96}$ & 0.75$_{\pm 0.02}$ \\
 &  & IUPAC Und. & 76.64$_{\pm 5.13}$ & 36.14$_{\pm 5.89}$ & 0.50$_{\pm 0.05}$ \\
 &  & CML Und. & 85.40$_{\pm 4.06}$ & 56.74$_{\pm 5.75}$ & 0.71$_{\pm 0.05}$ \\
 &  & Intra Norm. & 98.30$_{\pm 1.18}$ & 77.15$_{\pm 4.01}$ & 0.85$_{\pm 0.03}$ \\
 &  & Cross Norm. & 92.93$_{\pm 1.34}$ & 85.99$_{\pm 1.83}$ & 0.89$_{\pm 0.02}$ \\
\bottomrule
\end{tabular}
}

    \caption{MO conversion performance by task type for base models and the selected MolQwen3 text-only and multimodal models. InChIKey Match (IKM) is the primary metric.}
    \label{tab:conv_results}
\end{table}

\paragraph{Representation-level analysis.} Table~\ref{tab:input_output_reps} analyzes MO conversion performance by input and output molecular representation format, using InChIKey Match as the primary metric since it reflects structure-level correctness beyond exact string identity.

Among base-model configurations, randomized SELFIES to SELFIES attains the strongest performance, with $73.90\%$ IKM. Most other base-model configurations achieve near-zero IKM, indicating that zero-shot cross-representation conversion is limited. After post-training, most representation pairs reach $67\%$--$90\%$ IKM. SMILES to InChI reaches $100.00\%$ IKM, while randomized SMILES to SMILES reaches $95.60\%$. Randomized representations as inputs also achieve high IKM across several targets.

InChI remains difficult as an input representation: when InChI is the input, the best output IKM is $19.06\%$ for SMILES. In contrast, InChI is substantially easier as an output, with SMILES to InChI reaching $100.00\%$ and several other inputs reaching $85\%$--$94\%$. This asymmetry indicates that the learned mappings are direction dependent. One possible explanation is that generating a standardized InChI target is easier than extracting molecular structure from its layered input syntax; this interpretation would require further controlled analysis.

\begin{table}[tb]
    \centering
    \footnotesize
    \setlength{\tabcolsep}{2pt}

\resizebox{\linewidth}{!}{
\begin{tabular}{cl|cccc}
\toprule
 &  & \multicolumn{4}{c}{IKM (\%)} \\
 \cmidrule(lr){3-6}
 & Input\textbackslash Output Rep. & SMILES & SELFIES & DeepSMILES & InChI \\
%  &  Rep.  &  &  &  &  \\
\midrule
\multirow[c]{9}{*}{\rotatebox{90}{Base}} & SMILES & -- & 0.00$_{\pm 0.00}$ & 0.00$_{\pm 0.00}$ & 0.00$_{\pm 0.00}$ \\
 & SELFIES & 0.00$_{\pm 0.00}$ & -- & 0.00$_{\pm 0.00}$ & 0.00$_{\pm 0.00}$ \\
 & DeepSMILES & 2.22$_{\pm 2.21}$ & 0.00$_{\pm 0.00}$ & -- & 0.00$_{\pm 0.00}$ \\
 & InChI & 0.00$_{\pm 0.00}$ & 0.00$_{\pm 0.00}$ & 0.00$_{\pm 0.00}$ & -- \\
 & R. SMILES & 30.68$_{\pm 5.54}$ & 0.00$_{\pm 0.00}$ & 0.00$_{\pm 0.00}$ & 0.00$_{\pm 0.00}$ \\
 & R. SELFIES & 0.00$_{\pm 0.00}$ & \textbf{73.90}$_{\pm 4.69}$ & 0.00$_{\pm 0.00}$ & 0.00$_{\pm 0.00}$ \\
 & R. DeepSMILES & 0.00$_{\pm 0.00}$ & 0.00$_{\pm 0.00}$ & 13.77$_{\pm 4.18}$ & 0.00$_{\pm 0.00}$ \\
 & IUPAC & 0.00$_{\pm 0.00}$ & 0.00$_{\pm 0.00}$ & 0.00$_{\pm 0.00}$ & 0.00$_{\pm 0.00}$ \\
 & CML & 0.00$_{\pm 0.00}$ & 0.00$_{\pm 0.00}$ & 0.00$_{\pm 0.00}$ & 0.00$_{\pm 0.00}$ \\
\midrule
\multirow[c]{9}{*}{\rotatebox{90}{Trained}} & SMILES & -- & 85.94$_{\pm 5.79}$ & 79.53$_{\pm 6.92}$ & \textbf{100.00}$_{\pm 0.00}$ \\
 & SELFIES & 83.86$_{\pm 6.31}$ & -- & 90.12$_{\pm 4.32}$ & 87.34$_{\pm 4.78}$ \\
 & DeepSMILES & 86.16$_{\pm 5.09}$ & 81.62$_{\pm 5.91}$ & -- & 90.27$_{\pm 4.54}$ \\
 & InChI & 19.06$_{\pm 6.17}$ & 7.19$_{\pm 3.91}$ & 16.69$_{\pm 5.50}$ & -- \\
 & R. SMILES & 95.60$_{\pm 2.51}$ & 80.82$_{\pm 4.68}$ & 88.73$_{\pm 3.34}$ & 85.81$_{\pm 3.50}$ \\
 & R. SELFIES & 83.41$_{\pm 3.75}$ & 60.07$_{\pm 4.98}$ & 85.65$_{\pm 4.41}$ & 89.55$_{\pm 3.50}$ \\
 & R. DeepSMILES & 94.26$_{\pm 2.52}$ & 79.87$_{\pm 4.80}$ & 86.27$_{\pm 3.97}$ & 93.96$_{\pm 2.48}$ \\
 & IUPAC & 46.00$_{\pm 9.73}$ & 27.09$_{\pm 6.58}$ & 50.30$_{\pm 9.64}$ & 32.45$_{\pm 7.34}$ \\
 & CML & 55.84$_{\pm 8.35}$ & 53.04$_{\pm 8.45}$ & 66.63$_{\pm 6.87}$ & 47.59$_{\pm 8.37}$ \\
\bottomrule
\end{tabular}
}

    \caption{MO InChIKey Match (IKM) by input and output representation for base models and the selected MolQwen3 text-only and multimodal models. We report the average result across all instances.}
    \label{tab:input_output_reps}
\end{table}

\subsection{Rollout Analysis and Targeted Refinement}
\label{sec:rollout}

We ask whether refinement beyond supervised training helps, starting from the selected MQ3-VL-4B-I checkpoint. These experiments are diagnostic rather than part of the final recipe: we first analyze the effect of multiple sampling rollouts, then evaluate targeted SFT and GRPO on the same hard-case set.

Table~\ref{tab:rollout} reports pass@$k$ on the MO test set. Increasing the sample budget from $k=1$ to $k=8$ raises pass@$k$ from 67.89\% to 84.34\%, whereas the increment falls to 2.40 percentage points between $k=8$ and $k=16$. Thus, additional sampling recovers a subset of single-sample failures, but with diminishing returns.

The rollouts distinguish (i) \textit{easy} instances that pass at $k=1$ (67.89\%), (ii) \textit{recoverable} instances that fail at $k=1$ but pass by $k=8$ ($\sim$15\%), and (iii) instances not recovered at the evaluated budget of $k=16$ (13.26\%). Validity remains 87--88\% across $k$, so the additional recovered cases are not explained by a broad loss of chemical validity.

Both AC@$k$ (all correct) and AW@$k$ (all wrong) decrease as $k$ increases, showing heterogeneous sampled outcomes across instances. This descriptive result is not a calibration analysis.

\begin{table}[tb]
    \centering
    \footnotesize
    \setlength{\tabcolsep}{2pt}

\resizebox{\linewidth}{!}{
\begin{tabular}{c|ccccc}
\toprule
k & pass@k (\%) & $\Delta$ pass@k (\%) & valid@k (\%) & AC@k (\%) & AW@k (\%) \\
\midrule
1 & 67.89 & 67.89 & 87.09 & 67.89 & 32.11 \\
2 & 76.11 & 8.23 & 87.89 & 60.46 & 23.89 \\
4 & 80.11 & 4.00 & 87.89 & 54.06 & 19.89 \\
8 & 84.34 & 4.23 & 88.21 & 46.17 & 15.66 \\
16 & 86.74 & 2.40 & 88.29 & 39.09 & 13.26 \\
\bottomrule
\end{tabular}
}

    \caption{Rollout performance of MQ3-VL-4B-I on the MO test set. pass@k, valid@k, all-correct@k (AC@k), and all-wrong@k (AW@k) are reported for $k=1$ to $16$ generated samples per input.}
    \label{tab:rollout}
\end{table}

\paragraph{Targeted SFT.} We construct a 336-instance refinement set from the MO training data: recoverable instances that fail at pass@1 but pass by pass@8, plus small numbers of pass@1 successes and pass@8 failures. This mixture retains some easy and persistently unrecovered cases rather than optimizing only the recoverable subset.

Appendix~\ref{app:curves} shows the corresponding training curves. The targeted SFT checkpoint underperforms MQ3-VL-4B-I on the overall test set, indicating that this small refinement set does not provide a reliable improvement in this setting.

\paragraph{Targeted GRPO.} We also apply Group Relative Policy Optimization (GRPO)~\citep{shao2024deepseekmath} to the same refinement set, initialized from MQ3-VL-4B-I, using a ternary validity/InChIKey reward that directly optimizes chemical validity and exact structural recovery. For a prediction $y$ with reference InChIKey $y^\ast$, the sequence-level reward is
\[
r(y)=
\begin{cases}
-1, & y \text{ chemically invalid},\\
0, & y \text{ valid, InChIKey}(y)\neq y^\ast,\\
1, & y \text{ valid, InChIKey}(y)=y^\ast.
\end{cases}
\]
% We deliberately leave Tanimoto similarity out of the reward. A high TS does not imply progress toward the correct structure, since a molecule that shares most fingerprint bits with the reference can still be a different compound, and rewarding it risks teaching the model to produce plausible-looking neighbours rather than the target. The simple ternary reward therefore serves as a conservative sanity check: it credits only exact structure recovery.

We deliberately leave Tanimoto similarity out of the reward. A high TS does not imply progress toward the correct structure, rewarding it risks teaching the model to produce plausible-looking neighbours rather than the target.

Appendix~\ref{app:curves} compares GRPO training with MQ3-VL-4B-I. The reward fluctuates between steps but stabilizes over training relative to the supervised checkpoint. Under this data scale and ternary reward, GRPO therefore does not provide a reliable benefit over selected base checkpoints.

\paragraph{Decision.} Neither targeted SFT nor GRPO provides sufficiently consistent improvement over MQ3-VL-4B-I to justify inclusion in the final recipe. Under the current refinement-set size and reward design, both approaches can degrade broader test-set performance. We therefore report them as exploratory variants and retain the supervised checkpoints as the final models.

\begin{figure}[tb]
    \centering
    \includegraphics[width=\linewidth]{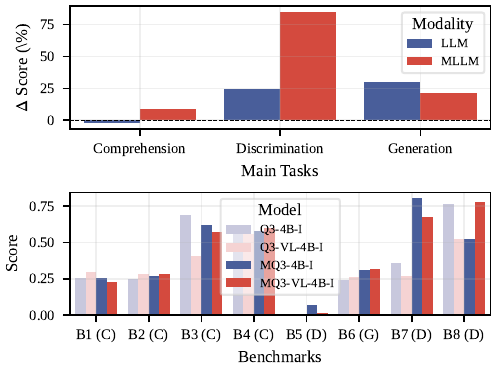}
    \caption{MolRepBench results for base models and MolQwen3 family. Results are grouped by the Comprehension, Discrimination, and Generation main task categories (top) and breakdown by each benchmark with which main task they belong to (bottom).}
    \label{fig:molrepbench}
\end{figure}

\subsection{External Task Transfer on MolRepBench}
\label{sec:transfer}

\paragraph{Protocol.} Table~\ref{tab:molrepbench_tasks} lists the eight MolRepBench tasks and their metrics. Unlike the conversion-focused MolReps benchmark, these tasks probe broader downstream molecular abilities, including atom counting, functional-group identification, property estimation, and structural-isomer discrimination, and thus require models to use information encoded in a representation beyond the conversion templates used for MolReps training.

\paragraph{Transfer results.} Figure~\ref{fig:molrepbench} compares the base models with the trained checkpoints. The post-trained models outperform their respective base models across the reported comprehension, discrimination, and generation aggregates, with an average relative improvement of 27.84\%. This pattern is consistent with representation-focused post-training improving performance beyond the MolReps conversion templates.

Improvements in comprehension are compatible with better extraction of structural information from molecular representations, while the discrimination and generation results show that gains are not restricted to direct format conversion. However, this evaluation does not isolate representation understanding as the sole causal mechanism: the models also receive additional molecular instruction data during post-training, and several MolRepBench tasks draw on ChEBI-20. The results should therefore be read as evidence of positive task transfer, rather than definitive evidence that improved representation understanding alone causes the downstream gains.

\begin{table}[tb]
\centering
\small

\resizebox{\linewidth}{!}{
\begin{tabular}{@{}l@{}llp{2.5cm}@{}}
\toprule
%\textbf{Task type} 
& \textbf{Task} & \textbf{Source} & \textbf{Metrics} \\
\midrule
\makebox[1.5ex][l]{Comprehension}\\
&Atom Counting (B1)         & ChEBI-20   & Exact match \\
&Group identification (B2)      & ChEBI-20   & Macro-F1 \\ % , 5 groups
%&Functional\ Group identification       & ChEBI-20   & Macro-F1 \\ % , 5 groups
&Property estimation (B3)  & ChEBI-20  & $\rho$, Exact match \\
&Molecule Retrieval (B4)    & ChEBI-20   & Top-1 \\
[1.5ex]
\makebox[1.5ex][l]{Discrimination}\\
&Isomer Discrimination (B5)      & ChEBI-20   & Accuracy \\
&Tautomer Recognition (B7)       & ZINC250K   & Accuracy \\
&Protonation State (B8)     & ZINC250K   & Accuracy \\
[1.5ex]
\makebox[1.5ex][l]{Generation}\\
&Caption-to-Molecule (B6)  & ChEBI-20   & Validity rate, \newline Tanimoto sim., \newline Exact match, FCD \raggedright \tabularnewline % Morgan fingerprint Tanimoto similarity
% Mol.\ Completion      & ChEBI-20 ($n$=200)  & Validity / Recovery \\
\bottomrule
\end{tabular}
}

    \caption{MolRepBench tasks used for external evaluation of molecular understanding from text representations. The benchmark comprises eight tasks in the Comprehension, Discrimination, and Generation categories.}
    \label{tab:molrepbench_tasks}
\end{table}

\section{Conclusion}
We introduced MolReps, a pipeline that probes whether language models relate the different representations of one molecule. Supervised post-training lifts cross-representation conversion from near zero to roughly 75\% InChIKey match when molecules are seen during training, but exact recovery on unseen molecules stays near 3\%, so the learned behavior is largely molecule-specific rather than general. Rollout-derived hard cases offered no reliable further gain under the current data scale. Because the same checkpoints improve on MolRepBench, format-level competence does transfer to other molecular tasks, even if structure-level generalization does not yet follow; scaling the pipeline with disjoint scaffolds is the natural next test.

\section*{Acknowledgement}
The authors thank Arun Raja for helpful discussions on experimental design and MolRepBench evaluation.

\bibliography{aaai2027}

% Check whether the conference requires a reproducibility checklist to be included in the paper.
% If so, you can uncomment the following line and ajust the path to include it.
% \input{ReproducibilityChecklist.tex}

\appendix

\section{MolReps Specification} \label{app:molreps}
The following sections give the full specification of the MolReps pipeline, complementing Section~\ref{sec:molreps}.

\subsection{Representations}\label{app:reps}
Table~\ref{tab:representations} lists the ten representations collected for each molecule, grouped into four families: linear notations, natural language, structured markup, and 2D drawings. Canonical SMILES, InChI, CML, and the 2D image are generated directly from the RDKit molecular graph, and the IUPAC name is retrieved from PubChem. SELFIES and DeepSMILES are converted from canonical SMILES. Their randomized variants are converted from randomized SMILES, so all three randomized forms describe the same non-canonical atom ordering.

\begin{table}[t]
    \centering
    \footnotesize
    \setlength{\tabcolsep}{4pt}

    \begin{tabular}{@{}lll@{}}
        \toprule
        Representation & Source & Tool \\
        \midrule
        \multicolumn{3}{@{}l}{\textit{1D: linear notations}} \\
        Canonical SMILES      & --            & \texttt{rdkit} \\
        Randomized SMILES     & canonical SMILES            & \texttt{rdkit} \\
        SELFIES               & canonical SMILES & \texttt{selfies} \\
        Randomized SELFIES    & rand.~SMILES     & \texttt{selfies} \\
        DeepSMILES            & canonical SMILES & \texttt{deepsmiles} \\
        Randomized DeepSMILES & rand.~SMILES     & \texttt{deepsmiles} \\
        InChI                 & canonical SMILES            & \texttt{rdkit} \\
        \addlinespace
        \multicolumn{3}{@{}l}{\textit{1D: natural language}} \\
        IUPAC name            & PubChem          & \texttt{pubchempy} \\
        \addlinespace
        \multicolumn{3}{@{}l}{\textit{1D: structured markup}} \\
        CML                   & canonical SMILES            & \texttt{rdkit} \\
        \addlinespace
        \multicolumn{3}{@{}l}{\textit{2D: visual}} \\
        2D image              & canonical SMILES            & \texttt{rdkit.Draw} \\
        \bottomrule
    \end{tabular}
    \caption{The ten representations collected for each molecule, the source each is generated from, and the generation tool. ``Graph'' denotes the RDKit molecular graph.}
    \label{tab:representations}
\end{table}

\subsection{Task Templates}\label{app:tasks}
Table~\ref{tab:tasks} gives the six task templates. Each template pairs an input representation with every admissible target in $\mathcal{R}_{\text{std}}$, and each pair is verbalized as a natural-language instruction. Normalization templates are instantiated once per randomized sample, so their count scales with $N$; with $N=2$ every molecule contributes 48 QA pairs.

\begin{table}[tb]
    \centering
    \small
    \setlength{\tabcolsep}{4pt}
    \begin{tabular}{@{}llr@{}}
        \toprule
        Task & Input $\to$ Output & \#/mol \\
        \midrule
        \multicolumn{3}{@{}l}{\textit{Cross-representation translation}} \\
        Std.~translation    & $r \to r'$;\ $r, r' \in \mathcal{R}_{\text{std}}$, $r' \neq r$ & 12 \\
        IUPAC understanding & IUPAC name $\to \mathcal{R}_{\text{std}}$ & 4 \\
        CML understanding   & CML $\to \mathcal{R}_{\text{std}}$ & 4 \\
        \addlinespace
        \multicolumn{3}{@{}l}{\textit{Normalization}} \\
        Intra-rep.~norm.    & $\mathcal{R}_{\text{rand}} \to$ same-family std. & $3N$ \\
        Cross-rep.~norm.    & $\mathcal{R}_{\text{rand}} \to$ 3 other-family std. & $9N$ \\
        \addlinespace
        \multicolumn{3}{@{}l}{\textit{Multimodal understanding}} \\
        OCSR                & 2D image $\to \mathcal{R}_{\text{std}}$ & 4 \\
        \midrule
        \multicolumn{2}{@{}l}{Total per molecule} & $24 + 12N$ \\
        \bottomrule
    \end{tabular}
    \caption{The six MolReps task templates. $\mathcal{R}_{\text{std}}$ = \{canonical SMILES, SELFIES, DeepSMILES, InChI\}; $\mathcal{R}_{\text{rand}}$ = \{randomized SMILES, SELFIES, DeepSMILES\}; $N$ is the number of randomized samples per molecule ($N=2$ in MolReps).}
    \label{tab:tasks}
\end{table}

\subsection{Dataset Statistics}\label{app:stats}
\begin{figure}[tb]
    \centering
    \includegraphics[width=0.85\linewidth]{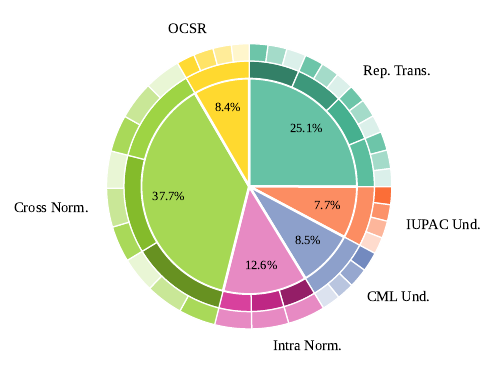}
    \caption{Distribution of task types over all 9,600 MolReps QA pairs. Rep.~Trans.: standardized representation translation; IUPAC Und.: IUPAC name understanding; CML Und.: CML understanding; Intra Norm.: intra-representation normalization; Cross Norm.: cross-representation normalization; OCSR: 2D structure recognition. See Section~\ref{sec:tasks} for details.} 
    \label{fig:task_distribution}
\end{figure}

Table~\ref{tab:statistics} reports the two evaluation protocols. The molecule-overlap (MO) split divides the 9,600 QA pairs into training, validation, and test sets at a ratio of 8:1:1, so test questions are new but their source molecules have been seen in training. The molecule-disjoint (MD) set is built from 20 further molecules that never appear in training and tests structural generalization. Figure~\ref{fig:task_distribution} shows the task-type distribution: the two normalization templates together account for about half of all pairs, because their count scales with the number of randomized samples.

\begin{table}[tb]
    \centering
    \small
        \begin{tabular}{lrr}
            \toprule
            Split & \# Src. Mol. & \# QA Pairs \\
            \midrule
            \multicolumn{3}{l}{\textit{QA-level split (molecule-overlap)}} \\
            Training set   & $\approx$200$^\dagger$ & 7680 \\
            Validation set & $\approx$200$^\dagger$ & 960 \\
            Test set       & $\approx$200$^\dagger$ & 960 \\
            \cmidrule(lr){2-3}
            \textit{~~Subtotal} & \textit{200} & \textit{9600} \\
            \midrule
            \multicolumn{3}{l}{\textit{Molecule-level hold-out (molecule-disjoint)}} \\
            Held-out test set & 20 & 960 \\
            \bottomrule
        \end{tabular}
    \caption{MolReps dataset splits. $^\dagger$The MO splits are disjoint at the QA-pair level but share approximately the same 200 source molecules; the MD held-out test set contains 20 molecules unseen during training.}
    \label{tab:statistics}
\end{table}

\section{Training Configuration}\label{app:training}
Table~\ref{tab:hyperparameters} lists the optimization settings. Both stages use LoRA adapters and are implemented with \texttt{trl}; GRPO rollouts are generated with \texttt{vLLM}. SFT is trained for at most ten epochs with early stopping on validation loss. GRPO samples a group of $G=8$ completions per prompt and scores each with the ternary reward of Section~\ref{sec:rollout}: $-1$ for an invalid output, $0$ for a valid but incorrect structure, and $1$ for an InChIKey match. We set the KL coefficient to zero, so the policy is regularized only by PPO-style ratio clipping.

\begin{table}[t]
    \centering
    \small

    \begin{tabular}{lcc}
        \toprule
        Hyperparameter & SFT & GRPO \\
        \midrule
        Learning rate & 1e-4 & 1e-5 \\
        Batch size & 16 & 32 \\
        Maximum epochs & 10 & 10 \\
        Optimizer & AdamW & AdamW \\
        KL coefficient $\beta$ & -- & 0 \\
        Group size $G$ & -- & 8 \\
        Clipping range $\epsilon$ & -- & 0.2 \\
        Reward & -- & $\{-1, 0, 1\}$ \\
        \bottomrule
    \end{tabular}

    \caption{Training hyperparameters for SFT and GRPO experiments.}
    \label{tab:hyperparameters}
\end{table}

\section{Training Curves}\label{app:curves}
\begin{figure}[tb]
    \centering
    \includegraphics[width=\linewidth]{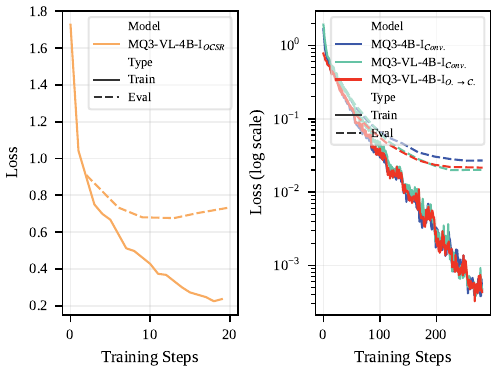}
    \caption{SFT training curves under different recipes on molecule OCSR task (left) and textual conversion task (right). Training and validation loss are tracked over training steps.}
    \label{fig:training_curves}
\end{figure}

\begin{figure}[tb]
    \centering
    \includegraphics[width=\linewidth]{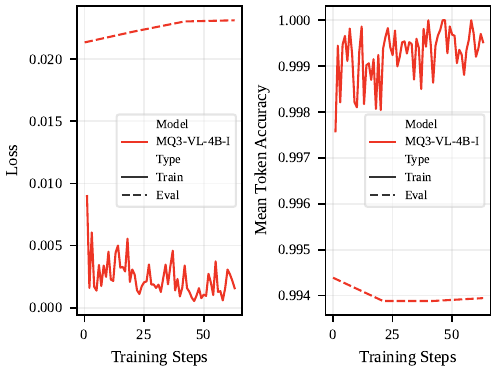}
    \caption{Targeted-SFT training curves of loss (left) and mean token accuracy (right). The 336-sample training set combines recoverable cases (fail at pass@1 and succeed by pass@8) with a small number of pass@1 successes and pass@8 failures.}
    \label{fig:sft_on_hard}
\end{figure}

\begin{figure}[tb]
    \centering
    \includegraphics[width=\linewidth]{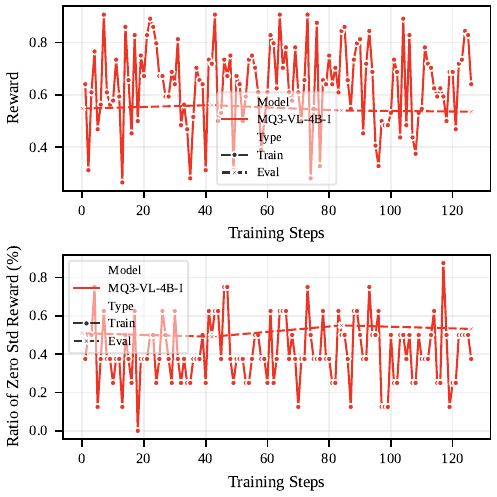}
    \caption{Targeted-GRPO training curves of reward (top) and ratio of zero-std reward in each batch (bottom). GRPO is initialized from MQ3-VL-4B-I and uses the ternary validity/InChIKey reward.}
    \label{fig:grpo_on_hard}
\end{figure}

Figure~\ref{fig:training_curves} shows the SFT curves for the post-training recipes compared in Section~\ref{sec:main}. Training and validation loss decrease together, so selecting the checkpoint with the lowest validation loss is reasonable. Figures~\ref{fig:sft_on_hard} and~\ref{fig:grpo_on_hard} show the targeted refinement runs of Section~\ref{sec:rollout}. Targeted SFT quickly fits the small refinement set. The GRPO reward fluctuates, and a large fraction of batches have zero reward variance, which leaves little learning signal.

\section{Examples of MolReps}
\label{app:examples}
This appendix illustrates the two levels at which MolReps is constructed. First, Table~\ref{tab:single-molecule-representations} shows the ten representations collected for a single molecule. The three randomized representations are represented by one sample each; multiple randomized serializations are generated in the dataset. Second, Table~\ref{tab:qa-examples} gives three resulting question--answer (QA) instances.

\subsection{Representations of a Single Molecule}
The example molecule is the anion with canonical SMILES \texttt{CC(C)CCCCCCCCCCC(=O)[O-]} (12-methyltridecanoate). It illustrates how the same molecular graph is expressed as linear notations, a chemical name, structured markup, and a 2D drawing.

\begin{table*}[tb]
    \centering
    \small

    \begin{tabular}{@{}p{0.2\textwidth}p{0.76\textwidth}@{}}
        \toprule
        Representation & Value \\
        \midrule
        Canonical SMILES & \texttt{\seqsplit{CC(C)CCCCCCCCCCC(=O)[O-]}} \\
        Randomized SMILES & \texttt{\seqsplit{CC(CCCCCCCCCCC([O-])=O)C}} \\
        SELFIES & \texttt{\seqsplit{[C][C][Branch1][C][C][C][C][C][C][C][C][C][C][C][C][C][=Branch1][C][=O][O-1]}} \\
        Randomized SELFIES & \texttt{\seqsplit{[C][C][Branch1][S][C][C][C][C][C][C][C][C][C][C][C][Branch1][C][O-1][=O][C]}} \\
        DeepSMILES & \texttt{\seqsplit{CCC)CCCCCCCCCCC=O)[O-]}} \\
        Randomized DeepSMILES & \texttt{\seqsplit{CCCCCCCCCCCCC[O-])=O))))))))))))C}} \\
        InChI & \texttt{\seqsplit{InChI=1S/C14H28O2/c1-13(2)11-9-7-5-3-4-6-8-10-12-14(15)16/h13H,3-12H2,1-2H3,(H,15,16)/p-1}} \\
        IUPAC name & 12-methyltridecanoate \\
        CML & \texttt{<molecule formalCharge="-1"> ... <atomArray> ... </atomArray> <bondArray> ... </bondArray> </molecule>} \\
        2D image & \raisebox{-0.5\height}{\includegraphics[height=5cm]{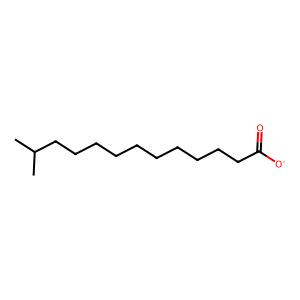}} \\
        \bottomrule
    \end{tabular}

    \caption{Ten MolReps representations for one molecule. A single sampled randomized form is displayed for each randomized representation; the CML document is abbreviated, and the 2D image is rendered with \texttt{rdkit.Draw}.}
    \label{tab:single-molecule-representations}
\end{table*}

\subsection{Question--Answer Examples}
Each instance supplies one representation in a natural-language instruction and requires an exact target representation. The examples below are cross-representation normalization tasks: a randomized DeepSMILES input is first interpreted and then converted to a canonical target representation.

\begin{table*}[tb]
    \centering
    \footnotesize
    \begin{tabular}{@{}p{0.23\textwidth}p{0.7\textwidth}@{}}
        \toprule
        Example & Content \\
        \midrule
        1: R.~DeepSMILES $\to$ SELFIES & \textbf{User:} Clean up the randomized DeepSMILES \texttt{\seqsplit{O=CCCCCCCCCCCCCCCC)))))))))))))))[O-]}} and give me the SELFIES. Just provide the SELFIES without any explanation. \\
        & \textbf{Assistant:} \texttt{\seqsplit{[C][C][C][C][C][C][C][C][C][C][C][C][C][C][C][C][=Branch1][C][=O][O-1]}} \\
        \midrule
        2: R.~DeepSMILES $\to$ SELFIES & \textbf{User:} Take \texttt{\seqsplit{[C@@]C)[C@H]CCC=C[C@H]6[C@@H]CC\%10))CC)C)))))C)))))O}} (a noisy DeepSMILES) and express it as proper SELFIES. Return only the SELFIES representation. \\
        & \textbf{Assistant:} \texttt{\seqsplit{[C][C][=C][C@H1][C@H1][Branch1][=Branch1][C][Branch1][C][C][C][C][C][C@][Branch1][C][C][Branch1][C][O][C@H1][Ring1][O][C][C][Ring1][\#C]}} \\
        \midrule
        3: R.~DeepSMILES $\to$ InChI & \textbf{User:} This DeepSMILES \texttt{\seqsplit{CO)[C@H]O[C@H]O)[C@H][C@@H]O[C@H]O[C@H]C)[C@H]O)C[C@H]6O))))))))[C@@H]6O)))O[C@@H][C@@H][C@H][C@H][C@@H]CO))O6))O))O))O}} may not be canonical. Normalize it and output as InChI. Please output the InChI representation only. \\
        & \textbf{Assistant:} \texttt{\seqsplit{InChI=1S/C18H32O14/c1-5-6(21)2-7(22)17(28-5)31-14-11(24)9(4-20)29-16(27)15(14)32-18-13(26)12(25)10(23)8(3-19)30-18/h5-27H,2-4H2,1H3/t5-,6-,7-,8-,9-,10+,11-,12+,13-,14+,15+,16+,17-,18-/m1/s1}} \\
        \bottomrule
    \end{tabular}
    \caption{Three MolReps QA examples, all cross-representation normalization tasks with a randomized DeepSMILES (R.~DeepSMILES) input. ``Assistant'' is the reference completion used for supervised post-training and evaluation.}
    \label{tab:qa-examples}
\end{table*}

\end{document}